\documentclass[prl,10pt,aps,twocolumn,superscriptaddress,showpacs,nofootinbib,floatfix,amssymb,amsmath]{revtex4-1}
\usepackage{rotating}

\usepackage{graphicx}
\usepackage{subfigure}
\usepackage{float}

\usepackage{amsmath, amsfonts, amssymb}
\newcommand{\be}{\begin{equation}}
\newcommand{\ee}{\end{equation}}
\newcommand{\bal}{\begin{align}}
\newcommand{\eal}{\end{align}}
\newcommand{\bmt}{\begin{multline}}
\newcommand{\emt}{\end{multline}}

\usepackage{color}

\newcommand{\ext}{{\mathrm{ext}}}

\newcommand{\beqn}{\begin{eqnarray}}
\newcommand{\eeqn}{\end{eqnarray}}
\newcommand{\eq}[1]{(\ref{#1})}

\newcommand{\W}{A}

\begin{document}

\title{On chromoelectric (super)conductivity of the Yang-Mills vacuum}

\author{M. N. Chernodub}\email{On leave from ITEP, Moscow, Russia.}
\affiliation{CNRS, Laboratoire de Math\'ematiques et Physique Th\'eorique, Universit\'e Fran\c{c}ois-Rabelais Tours,\\ F\'ed\'eration Denis Poisson, Parc de Grandmont, 37200 Tours, France}
\affiliation{Department of Physics and Astronomy, University of Gent, Krijgslaan 281, S9, B-9000 Gent, Belgium}
\author{Jos Van Doorsselaere}
\affiliation{CNRS, Laboratoire de Math\'ematiques et Physique Th\'eorique, Universit\'e Fran\c{c}ois-Rabelais Tours,\\ F\'ed\'eration Denis Poisson, Parc de Grandmont, 37200 Tours, France}
\affiliation{Department of Physics and Astronomy, University of Gent, Krijgslaan 281, S9, B-9000 Gent, Belgium}
\author{Tigran Kalaydzhyan}
\affiliation{DESY Hamburg, Theory Group, Notkestrasse 85, D22607 Hamburg, Germany}
\affiliation{Department of Physics and Astronomy, Stony Brook University, Stony Brook, New York 11794-3800, USA}
\author{Henri Verschelde}
\affiliation{Department of Physics and Astronomy, University of Gent, Krijgslaan 281, S9, B-9000 Gent, Belgium}

\begin{abstract}
We argue that in the Copenhagen (``spaghetti'') picture of the QCD vacuum the chromomagnetic flux tubes exhibit chromoelectric superconductivity. We show that the superconducting chromoelectric currents in the tubes may be induced by the topological charge density.
\end{abstract}


\date{\today}

\maketitle

The nonperturbative structure of the ground state of QCD vacuum is one of the most interesting unsolved problems in quantum field theory. At zero temperature  the ground state exhibits a mass gap, breaks chiral symmetry and supports confinement of color sources, quarks and gluons. The confining properties of the QCD ground state were intensively studied last decades resulting in a number of phenomenological approaches to this problem.

One of the popular approaches is the ``spaghetti vacuum'' picture (the Copenhagen vacuum): the QCD vacuum is considered to be populated by evolving vortex tubes which carry a chromomagnetic flux~\cite{ref:NO:1978:instability,ref:NO:electricstring,ref:NO:1979:spaghetti,ref:AO:lattice,ref:AO:center}. An isolated color charge -- for example, a quark --  scatters off the vortices and develops an infinitely large free energy. As a result, the quarks may appear in the vacuum only in a form of colourless (hadronic) states bounded by a chromoelectric string~\cite{ref:NO:electricstring}.

The standard mechanism of formation of the chromomagnetic vortices is as follows. The perturbative vacuum of QCD -- which is paramagnetic due to the asymptotic freedom -- has an unstable mode towards formation of a chromomagnetic field~\cite{ref:Savvidy}. However, in the background of a homogeneous chromomagnetic field the gluon part of the vacuum energy develops an imaginary part to large chromomagnetic moment of the gluon~\cite{ref:NO:1978:instability}. This implies that the homogeneous chromomagnetic field is also unstable towards squeezing of the chromomagnetic field into separate parallel flux tubes (vortices)~\cite{ref:AO:lattice}, similarly to the Abrikosov vortex lattice in a mixed state of an ordinary type-II superconductor in an external magnetic field~\cite{Abrikosov:1956sx}. Finally, due to global rotational and Lorentz  invariance of the QCD vacuum, the chromomagnetic field has locally a domain-like structure~\cite{ref:NO:1979:spaghetti}: the field has different orientation in different domains. Due to the fact that the vortices follow the orientation of the chromomagnetic field, the vortex lines form an intertwining entangled structure, hence the name ``spaghetti''.

Thus, the Copenhagen confining mechanism has a tight relation to ideas from ordinary superconductivity such as condensation and flux tube (vortex) formation. However, in addition to the mentioned features there exists another, primary phenomenon which is associated with ordinary superconductivity which is the superconductivity itself (i.e., the perfect conductivity of an electric current). In this paper we would like to show that the Copenhagen vacuum is not just ``analogous'' to an ordinary superconductivity: in this picture the Copenhagen vacuum {\emph{is}} a chromoelectric superconductor from the point of view of the transport properties.

Why the Copenhagen vacuum should be a chromoelectric superconductor? A simple answer is because in this picture the chromomagnetic tubes are formed due to the gluon condensate while the gluons are carrying a color charge. The condensation of the color charges should lead, naively, to the  (chromo)superconducting phenomenon. However, our considerations may contain a caveat: in the ordinary superconductivity the Cooper-pair condensate has a macroscopic order over large distances and this property is the core reason why the electric current may be transported by the uniform condensate without dissipation.  On the contrary, the QCD vacuum in the Copenhagen picture has a domain-like structure with each domain possessing its own orientation (both in color and coordinate spaces) of the gluon condensate so that the long-range order is absent. Nevertheless, we argue below that this property is not an obstacle due to the long-range order which is maintained {\emph{along}} the chromomagnetic vortices. We arrive to the picture that in the spaghetti vacuum the chromoelectric current should be able to stream without dissipation along the chromomagnetic tubes. Basically, the chromomagnetic tubes work like specific, hollow wires which are able to carry the chromoelectric current without resistance.

As one of the possible consequences of the color superconductivity one can expect probe quarks to propagate along the flux tubes over arbitrary distances, so that the tubes can be considered as ``fermionic guides'' \cite{Tiktopoulos:1986nc,Zhukovsky:2004ec,Kirilin:2012mw,Zakharov:2012vv}. From the phenomenological perspective the long-range propagation of quarks may lead to the phenomenon of chiral superfluidity of the quark-gluon plasma \cite{Kalaydzhyan:2012ut}.

The Yang-Mills Lagrangian is
\beqn
{\mathcal{L}} = - \frac{1}{4} F^{a}_{\mu\nu} F^{a\mu\nu}\,,
\label{eq:L}
\eeqn
where $F_{\mu\nu}^{a} = \partial_{\mu} \W^{a}_{\nu} - \partial_{\nu} \W^{a}_{\mu} + g \epsilon^{abc} \W^{b}_\mu\W^c_{\nu}$ is the strength tensor of the $SU(2)$ gluon field $\W^{a}_{\mu}$.

For simplicity, we consider the $SU(2)$ gauge field instead of more phenomenologically relevant $SU(3)$ fields since the latter solutions may be obtained -- following the general construction of Ambjorn--Olesen~\cite{ref:AO:lattice} -- by an imbedding the $SU(2)$ solutions into the $SU(3)$ color group.

The corresponding equations of motion are as follows
\beqn
\partial^\mu F_{\mu\nu}^a+g\epsilon^{abc}\W^{b\mu} F^c_{\mu\nu} = 0\,.
\label{eq:EOM:original}
\eeqn

Following Ambjorn and Olesen~\cite{ref:AO:lattice} we consider the state of the Yang-Mills theory in a uniform chromomagnetic field directed along the third spatial axis. In the color space the chromomagnetic field is assumed to be directed in the third axis as well:
\beqn
F^{a,\ext}_{\mu\nu} \sim \delta^{a,3} \left( \delta_{\mu1} \delta_{\nu2} - \delta_{\mu2} \delta_{\nu1} \right)\,.
\label{eq:external:field}
\eeqn
In order to obtain such a configuration, one can add a homogeneous abelian magnetic flux in third direction in color space~\cite{ref:AO:lattice}:
\beqn
\W^3_1=-x_2\frac{B}{2},\quad \W^3_2=x_1\frac{B}{2}\,. \label{eq:external:field:integrated}
\eeqn
For definiteness we take $B>0$.

The ground state solution to the equations of motion~\eq{eq:EOM:original} has certain remarkable properties. The solution is a function of the transverse -- with respect to the spatial direction of the external chromomagnetic field~\eq{eq:external:field} -- coordinates $x^{\perp} = (x_{1},x_{2})$ and it is independent on the longitudinal coordinates~$x^{\|} = (x_{0},x_{3})$.

The longitudinal components of the vector fields are vanishing in the ground state, $\W^{a}_{0} = \W^{a}_{3} = 0$, so that the equations of motion~\eq{eq:EOM:original} involve only the transversal components $\W^{a}_{i}$ with $i = 1,2$. The latter can conveniently be rewritten in the complex notations by introducing the following combinations for all vector fields $\mathcal{O}_{i}$ with $i=1,2$: $\mathcal{O}=\mathcal{O}_1+i\mathcal{O}_2$ and $\mathcal{\bar O}=\mathcal{O}_1-i\mathcal{O}_2$. These relations imply $\mathcal{\bar O} = \mathcal{O}^*$ for all real vector fields~$\mathcal{O}_{i}$. Defining the complex coordinate $z=x_1+ix_2$ and complex derivative $\partial=\partial_1+i\partial_2$, we find the non-canonical relations $\bar \partial z=\partial\bar z=2$ and $\partial z=\bar\partial\bar z=0$.

The off-diagonal gluonic fields $\W^{1,2}_{\mu}$ can be combined into two complex-valued fields:
\beqn
\W^{\pm}_{\mu} = \frac{1}{\sqrt{2}} \left( \W^{1}_{\mu} \mp i \W^{2}_{\mu} \right),
\eeqn
These combinations are not independent, $\W^{\pm}_{\mu} \equiv (\W^{\mp}_{\mu})^{\dagger}$, so that below we will work with the $\W^{-}_{\mu}$ field only.

The ground state can be described by two complex functions $\W = \W(x^{\perp})$ and $\W^{3} = \W^{3}(x^{\perp})$ with
\beqn
\W \equiv \W^{-} & = & \W^{-}_{1} + i \W^{-}_{2}\,,
\label{eq:W:polarization}\\
\W^{3} & = & \W^{3}_{1} + i \W^{3}_{2}\,,
\label{eq:W3:polarization}
\eeqn
and their complex conjugates. The combinations \eq{eq:W:polarization} and \eq{eq:W3:polarization} correspond to, respectively, the offdiagonal and diagonal components of the $\W^{a}_{i}$ fields. The color direction is defined by the background chromomagnetic field. The alternative (barred) combination of the off-diagonal $\W$ fields is zero in the ground state, ${\bar \W}^{-} = \W^{-}_{1} - i \W^{-}_{2} = 0$. Notice that ${\W^+} \equiv ({\bar \W}^-)^* = 0$ and ${\bar \W^+} = ({\W}^-)^*$.

The constraints (\ref{eq:external:field}) for $a=1,2$ can now be rewritten as a single complex equation:
\beqn
\bar\partial \W = - \frac{gB}{2} \bar z\, \W\,,
\label{eq:eom:1}
\label{one}
\eeqn
which is well known from the work of Abrikosov~\cite{Abrikosov:1956sx} to possess finite-energy solutions with a lattice symmetry.

The solution for this equation minimizing the remaining terms contributing to the energy integrated over the transversal plane
\beqn
E_\perp=\int \frac{1}{2}\left(F^3_{12}\right)^2 d^2 x_\perp=\int \frac{1}{2}\left(B-\frac{g}{2}\vert \W\vert^2\right)^2 d^2 x_\perp, \quad\label{eq:transversal:energy}
\eeqn
was constructed in terms of $\theta$-functions. In the background of the strong chromomagnetic field the vacuum structure resembles the Abrikosov lattice in the mixed phase of the type-II superconductors~\cite{Abrikosov:1956sx}. In analogy with the lattice of the Abrikosov vortices in a superconductor, the chromomagnetic field in Yang--Mills theory organizes itself in similar periodic structures~\cite{ref:AO:lattice}.

The ground state solution by Ambjorn and Olesen is given [up to a gauge factor due to a different parametrization of magnetic field~\eq{eq:external:field:integrated}] by the following formula~\cite{ref:AO:lattice}:
\beqn
 \W(x_1,x_2) & = & \phi_0\,e^{i g B x_2 \frac{x_1 + i x_2}{2}}\theta_3\left(\frac{(x_1 + i x_2) \nu}{L_B} , e^{\frac{2 i \pi}{3}}\right)\!, \qquad
 \label{eq:gluon:field} \label{liou}\\
 \nu & = & \frac{\sqrt[4]{3}}{\sqrt{2}}\,, \qquad L_B=\sqrt{\frac{2\pi}{gB}},
\eeqn
where $\theta_3$ is the third Jacobi theta function and the overall factor $\phi_0 \approx 2.9 \sqrt{B/g}$ is determined by minimization of the energy functional~(\ref{eq:transversal:energy}).

The global energy minimum is reached for the equilateral triangular lattice (which is also called the hexagonal lattice) solutions of Eq.~\eq{eq:eom:1}. Another local minimum is found for a square lattice.

The geometrical pattern of the lattice structure in the Yang--Mills theory is determined by the Abrikosov ratio,
	\beqn
	\beta_\mathrm{A}= \mathrm{Area}_\perp\left(\int dx_\perp^2 \vert \W\vert^4\right)/\left(\int dx_\perp^2\vert \W\vert^2\right)^2,
	\eeqn
which can be expressed in terms of generalized $\theta$-functions~\cite{ref:AO:lattice}. The global minimum of the energy functional~ (\ref{eq:transversal:energy}) is
\beqn
E_{\perp,\mathrm{min}}=\mathrm{Area}_\perp \frac{B^2}{2}\left(1-\frac{1}{\beta_\mathrm{A}}\right)\,,
\eeqn
where for the hexagonal structure the Abrikosov ratio is $\beta_\mathrm{A} \approx 1.16$ similarly to an ordinary type-II superconductor~\cite{ref:betaA:hexagonal}.

It is worth noting that even in models where the forth-order interaction terms are more complicated and, as a consequence, another definition of $\beta_\mathrm{A}$ is needed, one still finds that the global energy minimum still corresponds to the hexagonal lattice pattern~\cite{ref:betaA:hexagonal,ref:QCD:superconductivity,ref:EW:superconductivity,ref:Johanna}.

\begin{figure}[h]
		\begin{center}
			\includegraphics[scale=0.3]{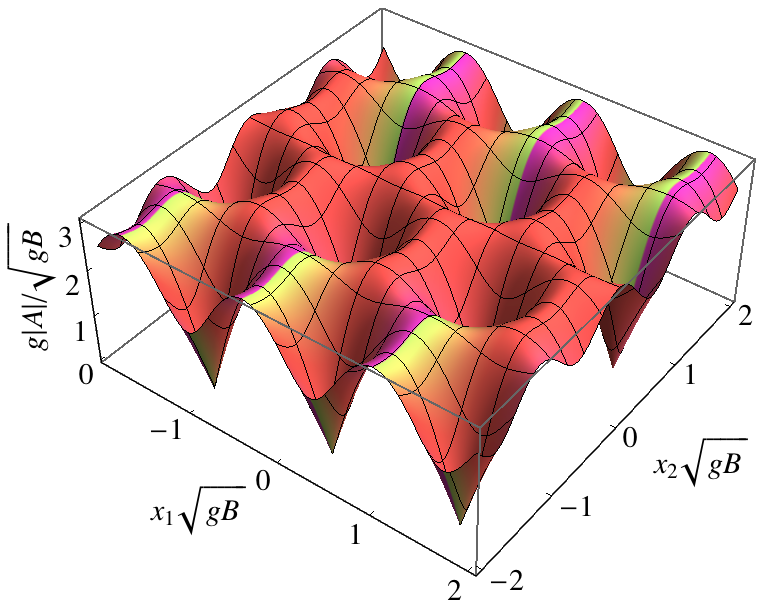}
		\end{center}
		\caption{The amplitude of the gluon field~\eq{liou} is shown by a density plot superimposed on the three dimensional plot of its absolute value. A few cells of the hexagonal periodic lattice are shown in the transverse $(x_1,x_2)$ plane. All dimensional units are shown in terms of the only massive scale $\sqrt{g B}$.}
\label{fig:gluon}
\end{figure}

The gluon field~\eq{eq:gluon:field} is shown in Fig.~\ref{fig:gluon}. The chromomagnetic vortices are arranged in the hexagonal structure. In the center of each vortex the gluon field~\eq{eq:W:polarization} is vanishing and the phase of this field winds by the angle $2\pi$, similarly to the usual Abrikosov vortex. The geometrical vortex pattern Fig.~\ref{fig:gluon} is identical to the Abrikosov vortex lattice in an ordinary type-II superconductor.

Does the ground state~\eq{liou} correspond to a chromoelectric superconductor? A simplest way to check an existence of the superconductivity is to calculate a relevant transport property: the superconducting nature of the ground state should reveal itself as an $\omega=0$ (zero-frequency) $\delta$-function peak in the real part of the complex (chromo)conductivity, $\sigma(\omega) = \sigma_1(\omega) + i \sigma_2(\omega)$. Alternatively, one can impose a weak external (chromo)electric field and then check that the ground state supports the London relation for the (chromo)electric currents. These two approaches are identical.

The London equations were used to argue in favor of existence of a magnetic-fields-induced electromagnetic superconductivity in QCD~\cite{ref:QCD:superconductivity} and in electroweak model~\cite{ref:EW:superconductivity}. The former effect is mediated by the $\rho$--meson condensation while the latter one is caused by the condensation of the $W$ mesons~\cite{ref:AO:EW}.

Following these approaches we impose a weak (test) chromoelectric field,
\beqn
E_3 \equiv E^3_3 = F^3_{30}\,, \qquad |E_3| \ll B\,,
\label{eq:E3}
\eeqn
oriented along the background chromomagnetic field~\eq{eq:external:field} both in the color and coordinate spaces in order to check possible validity of the London transport equation.

In order to define the relevant chromoelectric current we notice that the chromomagnetic field~\eq{eq:external:field} plays a role of an object which identifies the Abelian $U(1)$ direction in the non-Abelian $SU(2)$ group. The chromoelectric current associated with this $U(1)$ gauge subgroup is defined as follows\footnote{Notice that the chromoelectric current~\eq{eq:j:chromoelectric} is different from the full $SU(2)$ currents $\mathcal{D}^\mu F^a_{\mu\nu} \equiv 0$ since the Abelian and non-Abelian strengths are different: $ \partial_\mu \W^3_\nu - \partial_\nu \W^3_\mu \neq F^3_{\mu\nu}$.}:
\be
J_\nu \equiv J^3_\nu = \partial^\mu (\partial_\mu \W^3_\nu-\partial_\nu \W^3_\mu)\,.
\label{eq:j:chromoelectric}
\ee

We utilise the equations of motion~\eq{eq:EOM:original} in the background of the strong chromomagnetic~\eq{eq:external:field:integrated} and weak chromoelectric~\eq{eq:E3} fields and we find the following analogue of the London equation:
\beqn
\partial_{[0}J_{3]} =-g^2 \vert \W\vert^2 E_3\,,
\label{eq:London}
\eeqn
Equation~\eq{eq:London} implies, that the chromoelectric current propagates ballistically (i.e., without dissipation) along the chromomagnetic flux tubes. In the transverse directions the chromoelectric superconductivity is absent, $\partial_{[0}J_{i]} \equiv 0$ with $i=1,2$.

\begin{figure}[h]
\begin{center}
\includegraphics[scale=0.3]{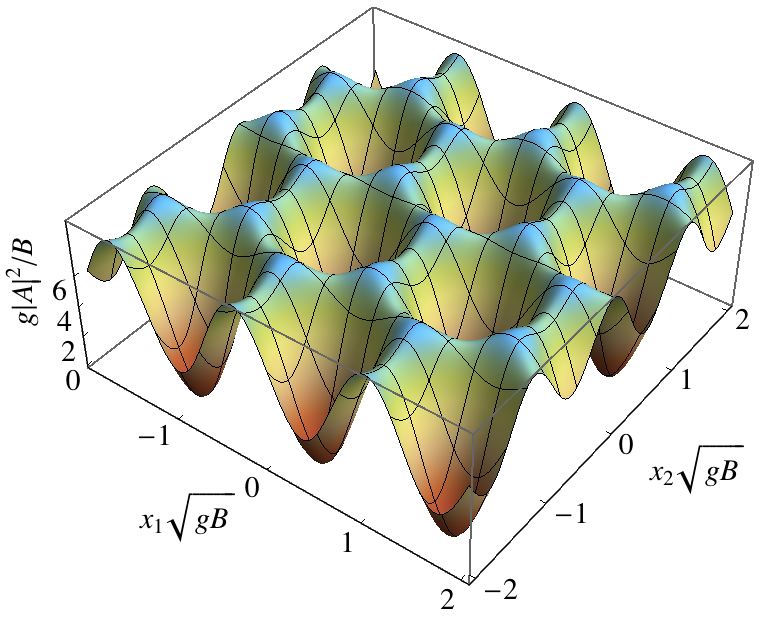}
\end{center}
\caption{The chromoelectric superconductivity coefficient in the London equation~\eq{eq:London} in the transverse $(x_1,x_2)$ plane.}
\label{fig:London}
\end{figure}

The superconductivity coefficient of the London equation~\eq{eq:London} is shown in Fig.~\ref{fig:London}. In the center of each flux tube the superconductivity is absent and the chromoelectric current may stream only at the regions in between the touching tubes. Therefore the chromomagnetic flux tube may be associated with a hollow (chromo)conducting ``wire''.

So far in our considerations we have followed the considerations of Refs.~\cite{ref:AO:lattice} where the regular solution was obtained in a homogeneous chromomagnetic background. In the real vacuum the flux tubes form an entangled spaghetti structure~\cite{ref:NO:1979:spaghetti}, so that the vacuum between two separated spatial points is, in general, disordered by the flux tubes. However, as we move along the tubes themselves they are supposed to keep their field structure in the transverse spatial directions~\cite{ref:NO:1979:spaghetti}. In lattice gauge theory the thick chromomagnetic flux tubes can be associated with the so-called center vortices~\cite{ref:Z:vortices}, which were indeed shown to exhibit the long-range correlations along the worldsheets of their flux tubes~\cite{ref:Zvortices:long:correlations}.

Interestingly, the chromoelectric currents~\eq{eq:London} are induced by the chromoelectric field $E^a_i$ provided it is parallel to the chromomagnetic field $B^a_i$ both in color and coordinate spaces. Thus, the electric currents in the flux tubes are induced if the scalar product $(\vec E^a \cdot \vec B^a)$ of these fields is nonzero. Notice, however, that this scalar product is proportional to the topological charge density~\cite{ref:instantons},
\beqn
q(x) = \frac{1}{16 \pi^2} {\mathrm {Tr}} \, [F_{\mu\nu} {\tilde F}^{\mu\nu}],  \qquad {\tilde F}^{\mu\nu} = \frac{1}{2} \varepsilon^{\mu\nu\alpha\beta} F_{\alpha\beta}. \quad
\label{eq:q}
\eeqn
Thus, the topological charge density should induce the chromoelectric currents in the chromomagnetic flux tubes. We find it fascinating that the described mechanism links chromoelectric superconductivity with the topology in QCD.

According to the standard Copenhagen picture the chromomagnetic tubes form an entangled ``spaghetti'' structure in the real vacuum. In our note we have shown that this spaghetti is (chromo)superconducting.

\vskip 5mm
\acknowledgments

The work of MCh was supported by Grant No. ANR-10-JCJC-0408 HYPERMAG (Agence nationale de la recherche, France).

\end{document}